\documentclass[conference]{IEEEtran}
\IEEEoverridecommandlockouts
\usepackage{cite}
\usepackage{amsmath,amssymb,amsfonts}
\usepackage{algorithmic}
\usepackage{graphicx}
\usepackage{textcomp}
\usepackage{xcolor}
\usepackage{comment}
\usepackage{multirow} 
\usepackage{caption}
\def\BibTeX{{\rm B\kern-.05em{\sc i\kern-.025em b}\kern-.08em
    T\kern-.1667em\lower.7ex\hbox{E}\kern-.125emX}}
\begin{document}

\title{Global and Local Attention-based Inception U-Net for Static IR Drop Prediction\\
\thanks{This work was partially supported by the Young and Middle-aged Teachers Education and Research Project of Fujian Province under grant JZ230001, the Natural Science Foundation of Fujian Province under Grant 2024J01363, the National Natural Science Foundation of China under Grant 92373207, and the National Key Research and Development Program of China under Grants 2022YFB4400500 and 2021YFA1003602.\\
*Corresponding author}
}

%\begin{comment}
\author{\IEEEauthorblockN{1\textsuperscript{st} Yilu Chen}
\IEEEauthorblockA{\textit{School of Computer and Information Engineering} \\
\textit{Xiamen University of Technology}\\
Xiamen, Fujian \\
yiluchenscholar@gmail.com}
\and
\IEEEauthorblockN{2\textsuperscript{nd} Zhijie Cai}
\IEEEauthorblockA{\textit{School of Microelectronics} \\
\textit{Fudan University}\\
Shanghai, China \\
zjcai22@m.fudan.edu.cn}
\and
\IEEEauthorblockN{3\textsuperscript{rd} Min Wei}
\IEEEauthorblockA{\textit{School of Microelectronics} \\
\textit{Fudan University}\\
Shanghai, China \\
mwei21@m.fudan.edu.cn}
\and
\IEEEauthorblockN{4\textsuperscript{th} Zhifeng Lin*}
\IEEEauthorblockA{\textit{Center for Discrete Mathematics and Theoretical Computer Science} \\
\textit{Fuzhou University}\\
Fuzhou, Fujian \\
linzhifeng@fzu.edu.cn}
\and
\IEEEauthorblockN{5\textsuperscript{th} Jianli Chen}
\IEEEauthorblockA{\textit{School of Microelectronics} \\
\textit{Fudan University}\\
Shanghai, China \\
chenjianli@fudan.edu.cn}
}
%\end{comment}

\maketitle

\begin{abstract}
Static IR drop analysis is a fundamental and critical task in chip design since the IR drop will significantly affect the design's functionality, performance, and reliability. However, the process of IR drop analysis can be time-consuming, potentially taking several hours. Therefore, a fast and accurate IR drop prediction is paramount for reducing the overall time invested in chip design. In this paper, we propose a global and local attention-based Inception U-Net for static IR drop prediction. Our U-Net incorporates the Transformer, CBAM, and Inception architectures to enhance its feature capture capability at different scales and improve the accuracy of predicted IR drop. Moreover, we propose 4 new features, which enhance our model with richer information. Finally, to balance the sampling probabilities across different regions in one design, we propose a series of novel data spatial adjustment techniques, with each batch randomly selecting one of them during training. Experimental results demonstrate that our proposed algorithm can achieve the best results among the winning teams of the ICCAD 2023 contest and the state-of-the-art algorithms.
\end{abstract}

\begin{IEEEkeywords}
IR drop prediction, U-Net, Attention, Transformer, Inception
\end{IEEEkeywords}

\section{Introduction}
\label{sec:introduction}
%TODO 要体现sign-off
Static Voltage (IR) drop analysis is a crucial task in sign-off, employed to assess the uniformity and stability of voltage distribution within a circuit. Non-uniform voltage distribution can potentially lead to performance degradation, increased power consumption, and circuit instability. 

Static IR drop refers to the voltage drops that occur in a circuit when it is in a steady-state condition. It is employed to identify weaknesses in the power delivery network (PDN) at the early design stages when switching vectors are unavailable. One of the main challenges of IR drop analysis is the extensive runtime of analysis tools, which may require several hours to complete an analysis for a single full-chip. Rapid and precise IR drop analysis enables engineers to accelerate the evaluation of various designs and promptly identify and rectify issues during the design phase without waiting for prolonged analysis results.

Traditional analysis methods involve the resolution of a large linear equation, denoted as $GV=J$ \cite{DBLP:conf/iccad/ZhongW05, DBLP:journals/fteda/ZhanKS08}, where $G$ represents the conductance matrix, $V$ represents the unknown voltage vector, and $J$ represents the current vector. Since this equation typically consists of millions to billions of variables, solving it may take several hours.

To accelerate the process, Kozhaya et al. \cite{DBLP:journals/tcad/KozhayaNN02} simplified the grid into coarser structures and then mapped the solutions back to the original grid to strike a balance between accuracy and speed. However, with the advancement of artificial intelligence technology, machine learning techniques have been regarded as superior approaches for IR drop prediction \cite{DBLP:conf/vts/LinFLLYLLF18, DBLP:conf/iccad/HoK19, DBLP:conf/aspdac/XieRKSSHC20, DBLP:conf/aspdac/ChhabriaAPPJS21, DBLP:conf/aspdac/ChenLWWLCLC22}, which can offer shorter estimation times and improved scalability compared to simulation-based methods. Ho et al. \cite{DBLP:conf/iccad/HoK19} proposed an IR drop predictor based on incremental analysis. However, their approach is not applicable for full-chip estimation. Xie et al. \cite{DBLP:conf/aspdac/XieRKSSHC20} proposed a full-chip IR drop prediction method, called PowerNet, based on convolutional neural networks (CNN). Nevertheless, their method requires measuring the total power for each grid, which entails dealing with multiple power maps. To address these problems, IREDGe \cite{DBLP:conf/aspdac/ChhabriaAPPJS21}, a full-chip U-Net \cite{DBLP:conf/miccai/RonnebergerFB15, oktay2018attention} technique that only involves processing a single power map, is presented. However, the features used by IREDGe might be insufficient, and the simple model might limit its potential to harness machine learning capabilities fully.

To address the aforementioned problems, in this paper, we propose a global and local attention-based Inception U-Net for static IR drop prediction. Our contributions can be summarized as follows:

\begin{itemize}
\item We propose an Inception-enhanced U-Net architecture to incorporate information from diverse scales to simultaneously consider the intrinsic characteristics of an instance and the surrounding information, thereby improving the accuracy in determining hotspots (high IR drop).
\item We propose a global attention block based on the Transformer \cite{DBLP:conf/nips/VaswaniSPUJGKP17, DBLP:conf/iclr/DosovitskiyB0WZ21} and a local attention block based on CBAM \cite{DBLP:conf/eccv/WooPLK18} to acquire a broader receptive field. The Transformer block is capable of capturing global information, while the CBAM block can gather wide-ranging local information.
\item We propose 4 new features, which gather richer information from the power delivery network. Enhanced by these features, our model could have a more comprehensive point of view and acquire better prediction precision.
\item We propose a series of novel data spatial adjustment techniques. Since random cropping or padding lead to a reduced probability of sampling at the edges, we devise 18 cropping or padding approaches to balance the sampling probabilities across different regions.
\item The experimental results demonstrate that our algorithm outperforms the winning teams of the ICCAD 2023 contest \cite{iccad_contest_2023} and the state-of-the-art methods.
\end{itemize}

The remainder of this paper is organized as follows. Section \ref{sec:preliminaries} presents the preliminaries of the problem. Section \ref{sec:algorithm} describes the details of the used features and our algorithm. Section \ref{sec:experiments} showcases the experimental results and ablative studies. Finally, Section \ref{sec:conclusion} concludes our algorithm.

\section{Preliminaries}
\label{sec:preliminaries}
\subsection{Problem Formulation}
\label{sec:problem_formulation}
The static IR drop relies on the following three factors: 1) the locations of voltage sources, 2) the topology of the PDN and the resistance values of each resistor in the network, and 3) the distribution of current sources.
In the static IR drop prediction problem of the ICCAD 2023 contest \cite{iccad_contest_2023}, the PDN resistance network is provided in the SPICE format, along with three image-based features, current map, PDN density map, and effective distance map, which are detailed in Section \ref{sec:features}. Based on the provided information, the objective is to predict the static IR drop with optimal precision.
\subsection{U-Net and Attention Mechanism}
U-Net \cite{DBLP:conf/miccai/RonnebergerFB15} is a classical image-to-image model, initially utilized in medical image segmentation. In the structure of U-Net, skip connections exist between the encoder and decoder, connecting the feature maps from the encoder directly to their corresponding layers in the decoder. This mechanism facilitates the preservation of more spatial information in segmentation tasks. Some U-Net variants enhance their performance by introducing new modules. Oktay et al. \cite{oktay2018attention} improved their model by incorporating an attention gate model to capture relevant information effectively. Punn et al. \cite{DBLP:journals/mta/PunnA21} enhanced their model's ability to capture features at different scales by combining U-Net with Inception \cite{DBLP:conf/cvpr/SzegedyVISW16, DBLP:conf/aaai/SzegedyIVA17}. However, they only integrated the basic version of Inception.

The attention mechanism is a crucial component in deep learning, enabling models to assign different weights to different parts of input data. However, traditional neural networks face challenges in achieving this. The convolutional block attention module (CBAM) \cite{DBLP:conf/eccv/WooPLK18} is an attention mechanism that can dynamically adjust the weights of spatial and channel information. Transformer \cite{DBLP:conf/iclr/DosovitskiyB0WZ21, DBLP:conf/iccv/LiuL00W0LG21, chen2021transunet} is a neural network architecture based on self-attention mechanisms, enabling efficient handling of long-range dependencies. By incorporating the Transformer, we propose an attention block capable of capturing global information.

\section{Algorithm}
\label{sec:algorithm}
In this section, we first introduce and visualize the features used in our model. Subsequently, we provide an overview of the architecture of our model. Then, the following sections provide detailed descriptions of each component of the model. The final section introduces the spatial adjustment techniques.

\subsection{Features Extraction}
% \section{Features and Ground Truth}
\label{sec:features}
\begin{figure}[h]
    \centering
    \includegraphics[width=1\linewidth]{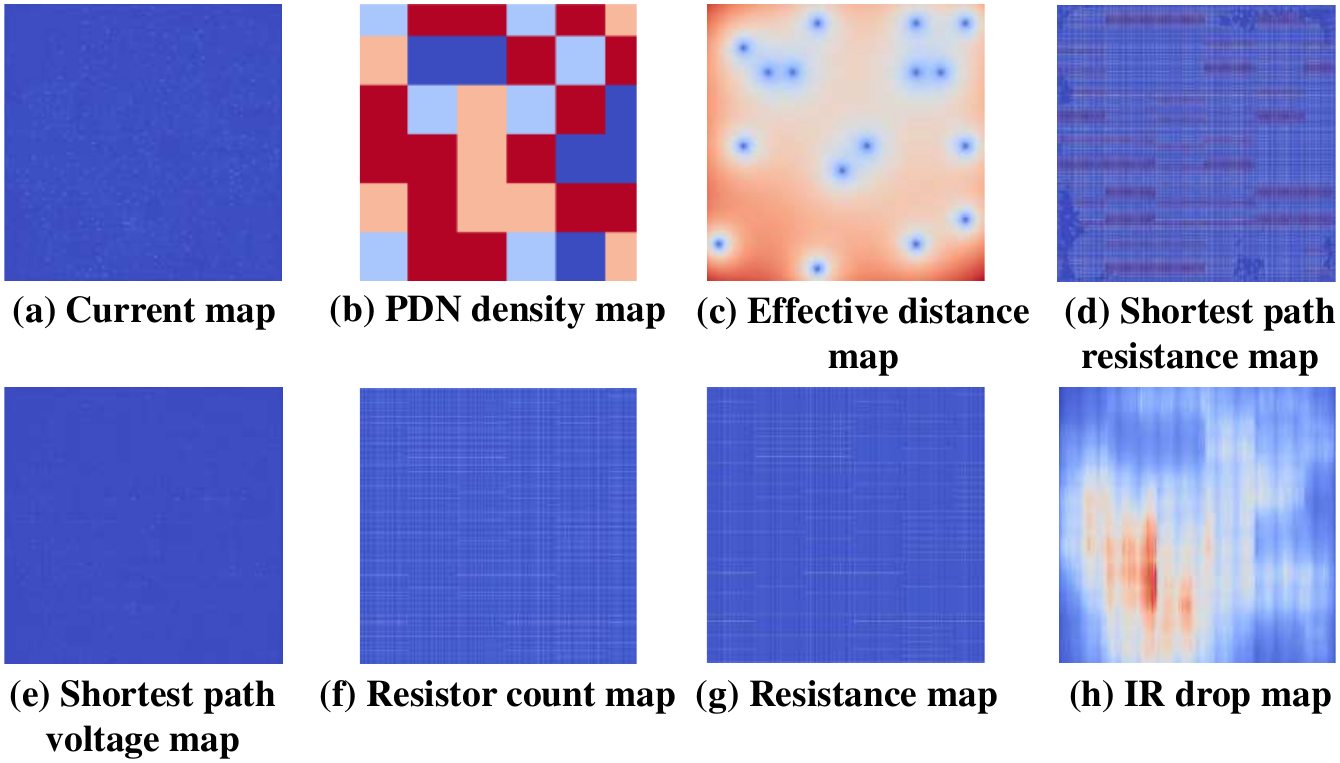}
    \caption{Visualization of seven used features and a ground-truth of IR drop map for testcase17. (a)(b)(c) are proposed in the literature \cite{DBLP:conf/aspdac/ChhabriaAPPJS21}. (d)(e)(f)(g) are our proposed new features.}
    \label{fig:features_and_label}
\end{figure}

\begin{figure*}[t]
    \centering
    \includegraphics[width=1\linewidth]{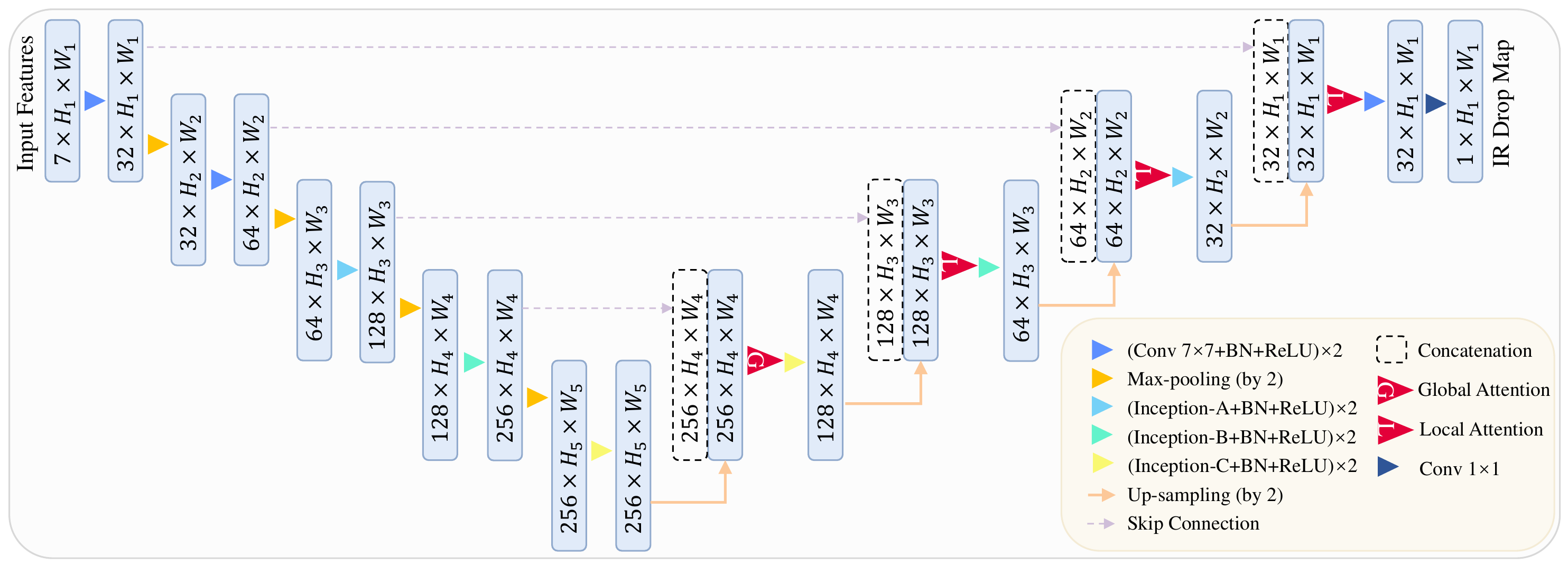}
    \caption{Our proposed model architecture, where $number \times H \times W$ signifies $channels \times height \times width$. During the encoding phase, the input features undergo four down-sampling operations with a factor of 2. Hence, $H_2 = H_1/2$, $H_3 = H_2/2$, and so forth. Conversely, during the decoding phase, four up-sampling operations are conducted with a factor of 2. We replace the majority of convolutions in the U-Net with Inception modules. To achieve a broader perceptual field, at the beginning of the decoder, we employ a global attention block, primarily utilizing the Transformer architecture. In the subsequent stages, we integrate the local attention blocks, whose pillar is CBAM. The schematics of these two attention blocks are respectively depicted in Figure \ref{fig:global_atten} and Figure \ref{fig:local_atten}.}
    \label{fig:architecture}
\end{figure*} 

Feature selection plays a pivotal role in machine learning, exerting a profound influence on the performance and efficiency of models. In this paper, we employ 3 features proposed in the literature \cite{DBLP:conf/aspdac/ChhabriaAPPJS21}, namely, the current, effective distance, and PDN density maps, as illustrated in Figure \ref{fig:features_and_label}(a)(b)(c), respectively. The current map depicts the distribution of current sources across the design. The effective distance, which provides an evaluation of the distance from each location to all voltage sources, is calculated as the reciprocal of the sum of the reciprocals of the Euclidean distances. Namely, $d_e = \left(d_1^{-1}+d_2^{-1}+...+d_K^{-1}\right)^{-1}$, where $d_e$ is the effective distance and $d_i$ is the Euclidean distance of the instance to the $i^{th}$ power source. Besides, the PDN density map is constructed by extracting the average PDN pitch within each grid. Furthermore, we propose 4 new features, which we categorize into two classes: instance-level features and PDN-structure-level features.
%In the context of IR drop prediction, we have devised two distinct categories of features: instance-level and PDN-structure-level, each providing valuable global and local information.
To generate the image feature, we initially partition the entire chip region into a 2D mesh, wherein each grid within the mesh maintains a fixed size of $1\mu m \times 1\mu m$.

%TODO 1、解释一下怎么去掉优先队列，2、power pad
As for the instance-level features, we incorporate the shortest path resistance map and shortest path voltage map, as depicted in Figure \ref{fig:features_and_label}(d)(e). Given that the IR drop of each instance relies on the resistance paths and the current flowing through them, we utilize the shortest path resistance and voltage maps to strike a balance between the relevancy and efficiency of feature extraction. The shortest path resistance is the cumulative resistance encountered along the shortest path from each instance to the nearest voltage source via the shortest path algorithm. To minimize computation time, we initiate calculations from all power sources, determining the shortest path to each instance. This approach necessitates executing the shortest path algorithm only $K$ times, where $K$ is the number of power sources, thereby enhancing efficiency. Hence, the total time complexity is $O\left(K\times \left(I\times\log_2{I}+E\right)\right)$,  where $I$ is the number of instances and $E$ is the number of edges. Since the PDN structure is quite sparse (an instance connects to at most 6 edges), we do not need to maintain a priority queue and $E<6\times I \div 2 = 3 \times I$. Therefore, the optimized time complexity can be $O\left(K\times \left(6\times I+3\times I\right)\right) = O\left(9\times K\times I\right)$. Specially, some instances may reside in the same grid. Hence, we generate the shortest path resistance map by computing the average value of the shortest path resistances for all instances within the same grid. Moreover, the shortest path voltage is derived by multiplying the shortest path resistance with the instance current.

As for the PDN-structure-level features, we incorporate the resistor count map and resistance map, as depicted in Figure \ref{fig:features_and_label}(f)(g). The resistance map and resistor count map count the resistance information for each grid, which only requires traversing each resistor one time. So, their time complexity is $O(WN)$, where $W$ is the maximum grid number of an edge, and $N$ is the number of resistors. We introduce the resistor count map and the resistance map to encompass more details concerning the PDN structure. Specifically, the resistor count map indicates the number of resistors passing through each grid, while the resistance map is computed by evenly distributing the resistance of each resistor to the corresponding overlapped grids. It is worth noting that we project the 3D PDN onto a 2D panel to obtain the 2D feature maps and mitigate the sparsity of features.

%Figure \ref{fig:features_and_label}(d)(e)(f)(g) depict the 4 new features employed for our model and the IR drop ground truth map of testcase17. Note that the current map, effective distance map, and PDN density map are proposed by \cite{DBLP:conf/aspdac/ChhabriaAPPJS21} and provided as default input features in ICCAD 2023 contest \cite{iccad_contest_2023}
%, while the other features are our proposed new features.

\subsection{Overall Architecture}

\begin{figure*}[t]
    \centering
    \includegraphics[width=1\linewidth]{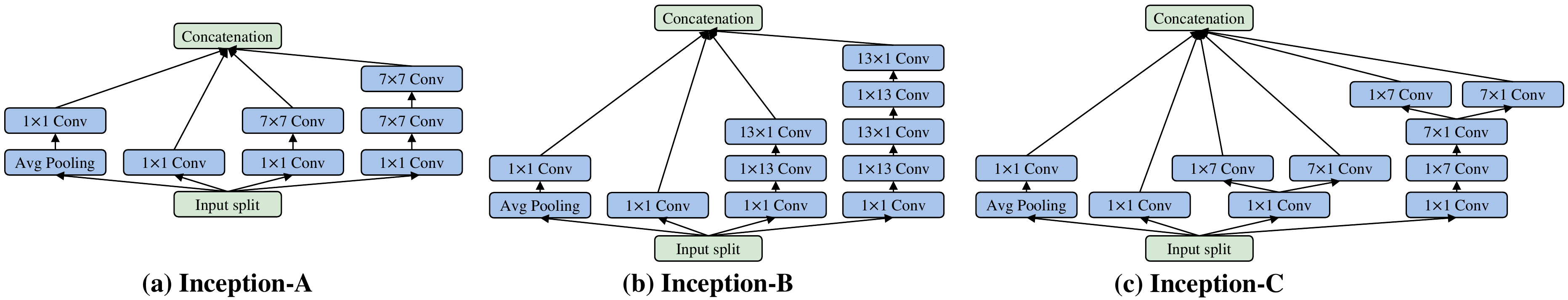}
    \caption{Three different Inception modules.}
    \label{fig:inception}
\end{figure*}

Our proposed model architecture is depicted in Figure \ref{fig:architecture}. The architectural backbone is based on U-Net \cite{DBLP:conf/miccai/RonnebergerFB15, oktay2018attention}. The input features undergo four down-sampling operations by a factor of 2 during the encoding phase. Hence, $H_2 = H_1/2$, $H_3 = H_2/2$, and so forth. Conversely, during the decoding phase, four up-sampling operations by a factor of 2, each employing deconvolution, are performed to restore the spatial dimensions. Following \cite{DBLP:conf/cvpr/0003MWFDX22}, since larger kernel sizes imply better performance, we replace all convolution kernels with larger kernel sizes. To capture low-level features, we start with a sequence of two $((7\times 7 \text{convolution} + \text{BN} + \text{ReLU})\times 2 + \text{Max-pooling})$ operations, where BN represents the batch normalization. Subsequently, to enhance the network's ability to combine local details and surrounding information, we incorporate the Inception modules \cite{DBLP:conf/cvpr/SzegedyVISW16, DBLP:conf/aaai/SzegedyIVA17}.

Based on our observations, it has become evident that the hotspot region of IR drop is not solely influenced by the local information. In fact, the surrounding environment at a larger scale can also exert a noticeable impact. Therefore, the necessity of an attention mechanism with a broader field of view has arisen. Consequently, we propose two key attention modules: a global attention block at the beginning of the decoder and a local attention in the subsequent stages. Finally, the prediction of the IR drop map is derived through an $1\times 1$ convolution operation. In the following sections, we will provide a comprehensive exposition of the aforementioned modules and blocks.

\subsection{Inception}

Inception modules \cite{DBLP:conf/cvpr/SzegedyLJSRAEVR15, DBLP:conf/cvpr/SzegedyVISW16, DBLP:conf/aaai/SzegedyIVA17} are a multi-branch convolutional architecture that allows the model to simultaneously learn feature maps from different convolutional kernel sizes. Such a multi-branch architecture assists the model in capturing features at various scales, thereby enhancing its performance. Inception modules come in three different variations \cite{DBLP:conf/cvpr/SzegedyVISW16, DBLP:conf/aaai/SzegedyIVA17}, as illustrated in Figure \ref{fig:inception}, namely, Inception-A, Inception-B, and Inception-C. As mentioned in \cite{DBLP:conf/cvpr/SzegedyVISW16}, the Inception-C module primarily focuses on high-dimensional feature extraction, and, in \cite{DBLP:conf/cvpr/SzegedyVISW16}, the sequence of Inceptions in the model is Inception-A, Inception-B, and Inception-C. Hence, we employ Inception-A, Inception-B, and Inception-C sequentially in the encoding phase. This arrangement maintains symmetry in the decoding phase. The kernel sizes of Inception modules used in our model architecture are annotated in Figure \ref{fig:inception}. 

\subsection{Global Attention}
\begin{figure}[h]
    \centering
    \includegraphics[width=1\linewidth]{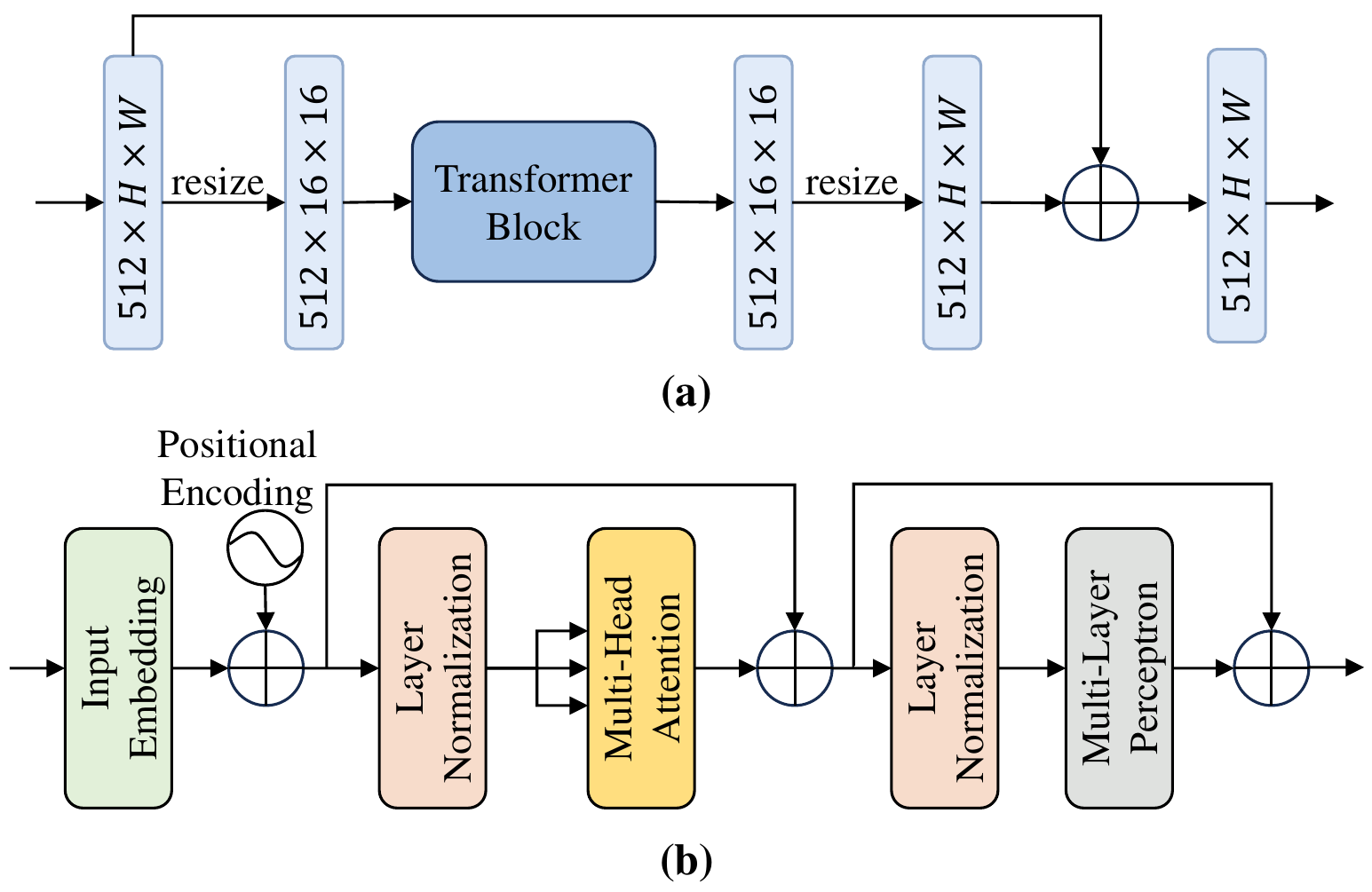}
    \caption{Global attention block. (a) is the overall flow of the global attention block. (b) is the diagram of the Transformer block, where $\oplus$ represents addition.}
    \label{fig:global_atten} 
\end{figure}

Figure \ref{fig:global_atten} depicts the global attention block. This block is a residual attention module based on the Transformer architecture \cite{DBLP:conf/iclr/DosovitskiyB0WZ21, DBLP:conf/iccv/LiuL00W0LG21, chen2021transunet}. In the Transformer, global information perception is achieved through three parameters, known as the Query, Key, and Value. Previous Transformer models faced limitations, as they could only accept fixed-sized inputs or required interpolating positional encoding to match the input size. To conserve computational resources, our proposed algorithm resizes features to a fixed $16\times 16$ dimension and then resizes the output of the Transformer block back to the original size $H\times W$, where the resizing technique is a bicubic interpolation. In this paper, the hidden size of the Transformer is set to match the number of channels in the features, i.e., $512$. The global information obtained from the Transformer is introduced to the original features through a residual addition mechanism. 

\subsection{Local Attention}

\begin{figure}[h]
    \centering
    \includegraphics[width=1\linewidth]{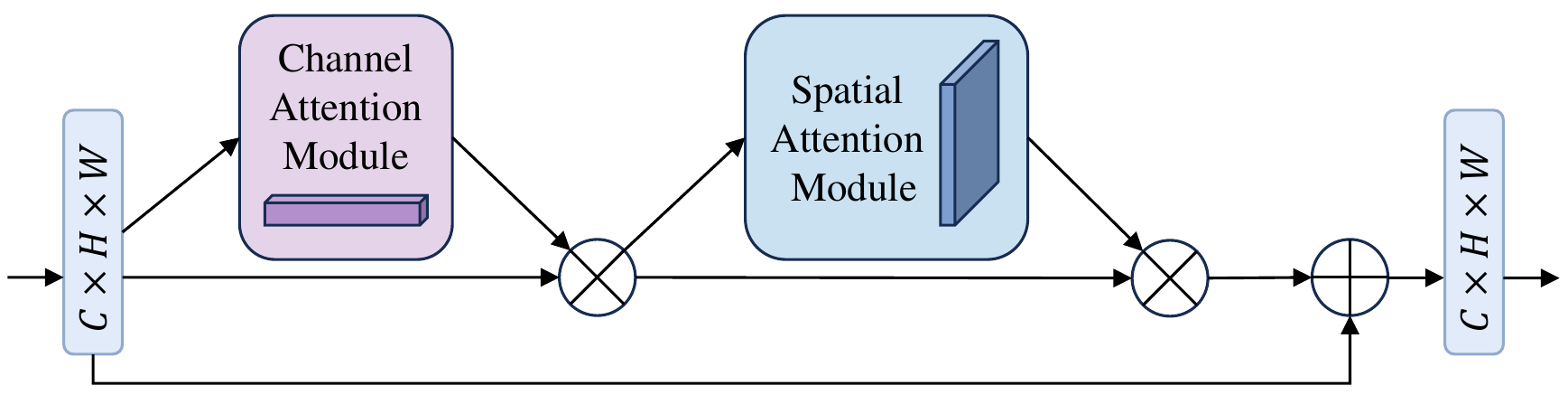}
    \caption{Local attention block, where $\oplus$ represents addition and $\otimes$ represents multiplication. For detailed architectural specifics, please refer to \cite{DBLP:conf/eccv/WooPLK18}.}
    \label{fig:local_atten}
\end{figure}

Due to the substantial computational demands of the Transformer and the potential loss of critical information when resizing features to $16\times 16$, we employ the CBAM \cite{DBLP:conf/eccv/WooPLK18}, as depicted in Figure \ref{fig:local_atten}, in the subsequent modules to implement local but not narrow attention. CBAM's spatial attention is realized by applying large-sized convolution kernels to the spatial features extracted from max-pooling and average-pooling. In this paper, we set this kernel size as $25$. 

\subsection{Data Spatial Adjustment Techniques}
\begin{figure}[h]
    \centering
    \includegraphics[width=1\linewidth]{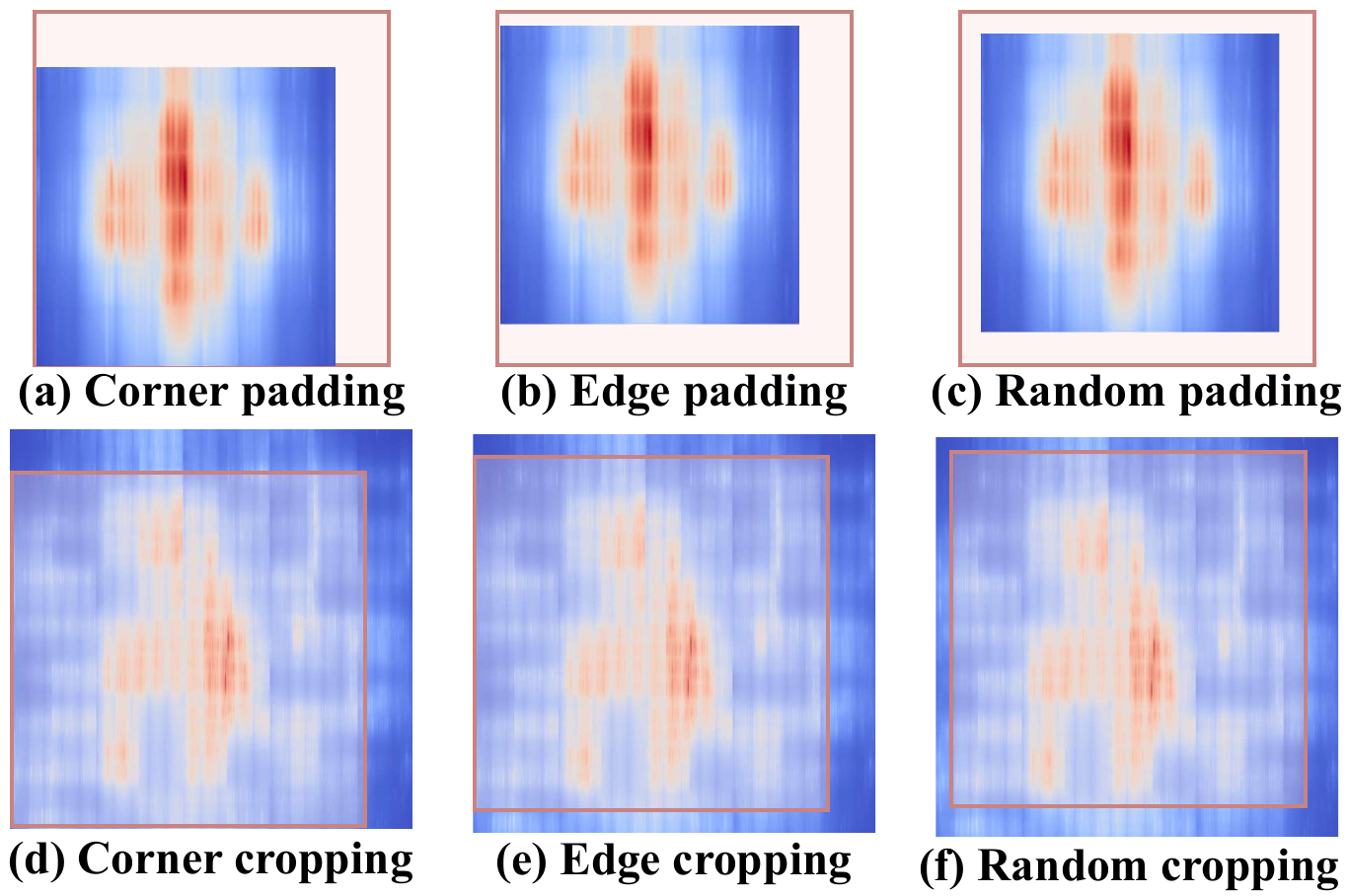}
    \caption{Several examples of spatial adjustment techniques.}
    \label{fig:crop_sample} %cropping or padding approaches
\end{figure}

The spatial dimensions (width$\times$height) of cases in the dataset vary significantly, with edge lengths ranging from 201 to 930. However, each batch requires maintaining consistent spatial dimensions in model training. Therefore, spatial adjustment techniques are necessary. Given the spatial dimensions of a batch as $l\times l$, a fundamental principle is that if the edge length is less than $l$, padding with zeros is performed. In contrast, if the edge length exceeds $l$, cropping is executed to make the edge length match $l$. However, if cropping or padding is randomized, it would result in a disproportionately lower probability of capturing edge regions compared to other areas. On the contrary, the probability of occurrence of the central region is maximized since almost every cropped area will include the central region. In light of this, we propose several distinct spatial adjustment techniques, whose examples are illustrated in Figure \ref{fig:crop_sample}. 

The spatial adjustment techniques encompass two primary categories, namely padding and cropping, both of which are employed to regulate the spatial dimensions. Further classification is based on the relative position, resulting in three distinct types: corner, edge, and random.
In the corner and edge types, the original input aligns with the target region in the corresponding corner or along the side edge. Conversely, the random type adopts a randomized approach to determine the relative position between the original and target regions. Considering each distinct edge and corner as separate methods, there are a total of 18 cropping or padding methods.

During model training, each batch is subjected to a random selection of one of these methods for cropping or padding. However, only when the edge length is less than $l$, padding will be applied; conversely, if it exceeds $l$, cropping will be implemented. These spatial adjustment techniques significantly increase the occurrence probability of edge regions and ensure diversity in the cropping and padding approaches, which enhances the overall performance and robustness of our model.

\section{Experiments}
\label{sec:experiments}
\subsection{Dataset}
The ICCAD 2023 contest \cite{iccad_contest_2023} provides 100 fake cases and 10 real cases for training, with additional 10 real cases designed for testing. Due to the limited training data, we augment our dataset with 2000 cases from nangate45\_set1 and nangate45\_set2 provided by BeGAN \cite{DBLP:conf/iccad/ChhabriaKZS21}. We select 5 real cases as a validation set, while the remaining cases constitute the training set. Furthermore, as the provided data from the ICCAD 2023 contest closely resembles the distribution of real cases in the testing stage, we oversample each fake case 10 times and each real case 20 times. Consequently, the training set comprises a total of 3100 cases. Our model will save the weights corresponding to the best performance on the validation set.
\subsection{Experimental Setup}
Our algorithm is implemented in Python with Pytorch. The training phase of the model is conducted on a 64-bit Ubuntu workstation with one NVIDIA A40 containing 48GB of GPU memory. To ensure runtime comparability, the testing phase is performed on Ubuntu 20.04 with a 16GB 2.42GHz Intel Core.
%The experiments are conducted on a 64-bit Ubuntu 20.04 with a 2.42GHz Intel Core and one NVIDIA A40 containing 48GB of GPU memory. 
%To ensure runtime comparability, the testing phase is performed on Ubuntu 20.04 with a 16GB 2.42GHz Intel Core.
%TODO 训练时间, VDD是多大？
Each batch randomly selects an edge length $l$ from the range of 496 to 512. We use the RMSprop as the optimizer with a batch size of $32$, a momentum of $0.9$, and a weight decay of $10^{-8}$. The initial learning rate is set at $0.001$. The loss function is the mean squared error (MSE). The total number of epochs is 600. There is a 50-epoch linear warmup and a cosine decaying schedule afterward. In addition to the 18 cropping or padding approaches, we also employ random horizontal and vertical flips for data augmentation. The standard deviations of the features and ground truth are computed, and they are used to normalize the data by dividing each of them by their respective standard deviations. We do not subtract the mean of data in order to preserve the zero values. The total training time is 10 hours. The VDD of all cases is 1.1V.
\subsection{Evaluation Metrics}
%TODO 增加error的定义
We follow the ICCAD 2023 contest in selecting the mean absolute error (MAE), the worst-case IR drop-based F1 score, and runtime as the evaluation metrics. The MAE is computed using the following formula.
\begin{equation}\label{equation:mae}
   MAE = \frac{1}{N_{points}}\sum^{N_{points}}_{i=1}\left| y_i-x_i\right|,
\end{equation}
where $N_{points}$ represents the number of points on the IR drop map, $y_i$ denotes the ground truth at the $i^{th}$ position and $x_i$ represents the predicted value at the $i^{th}$ position.

Furthermore, since sign-off is to ensure the robustness of the power grid, the worst-case IR drop is important in the IR drop analysis. The F1 score serves as an indicator of the predictive performance for the worst-case IR drop. In this problem, IR drop values exceeding 90\% of the maximum ground truth are classified as positive samples, while the rest are classified as negative samples. The predicted results are categorized into four groups, i.e., true positive (TP), true negative (TN), false positive (FP), and false negative (FN). TP/TN represent correctly predicted positive/negative samples, while FN/FP denote incorrectly predicted positive/negative samples. Hence, the formula of F1 is as follows.

\begin{align}\label{equation:F1}
  P &= \frac{TP}{TP+FP}, \\
  R &= \frac{TP}{TP+FN}, \\
  F1 &=  \frac{2\times P \times R}{P+R}.
\end{align}

Furthermore, to further analyze the worst-case IR drop in sign-off, we also present the errors associated with the worst-case IR drop ($W_{IR}$) and the maximum error. The error is defined as $V_{err}=|V_{true}-V_{pred}|$, where $V_{true}$ represents the actual IR drop value, and $V_{pred}$ denotes the predicted IR drop at the corresponding position. Specifically, the worst-case IR drop error refers to the error between the worst-case IR drop in the ground truth and the corresponding predicted IR drop. Moreover, the maximum IR drop error (Max $IR_{err}$) means the largest error across all errors.

\subsection{Experimental Results}
\begin{table*}[!t]
\caption{Experimental results of the compared algorithms and our algorithm, where the unit of MAE is mV.}
\begin{tabular}{|c|ccc|ccc|ccc|ccc|ccc|}
\hline
\multirow{2}{*}{Testcase} & \multicolumn{3}{c|}{1st Place} & \multicolumn{3}{c|}{2nd Place} & \multicolumn{3}{c|}{3rd Place} & \multicolumn{3}{c|}{IREDGe \cite{DBLP:conf/aspdac/ChhabriaAPPJS21}} & \multicolumn{3}{c|}{Ours} \\ \cline{2-16} 
                  & MAE     & F1     & Time(s)     & MAE     & F1      & Time(s)    & MAE     & F1     & Time(s)     & MAE       & F1       & Time(s)     & MAE    & F1    & Time(s)  \\ \hline
\hline

testcase7                  & 0.066    & 0.78    & 14.61      & 0.078    & 0.56    & 3.22       & 0.075    & 0.87    & -          & 0.505      & 0.14     & 1.42        & {\color[HTML]{333333} 0.081} & {\color[HTML]{333333} 0.69} & {\color[HTML]{333333} 3.33} \\
testcase8                  & 0.082    & 0.82    & 12.64      & 0.113    & 0.80    & 2.70       & 0.114    & 0.87    & -          & 0.479      & 0.21     & 1.37        & {\color[HTML]{333333} 0.104} & {\color[HTML]{333333} 0.63} & {\color[HTML]{333333} 3.08} \\
testcase9                  & 0.041    & 0.59    & 18.84      & 0.073    & 0.55    & 4.25       & 0.051    & 0.47    & -          & 0.480      & 0.02     & 2.30        & {\color[HTML]{333333} 0.057} & {\color[HTML]{333333} 0.72} & {\color[HTML]{333333} 4.19} \\
testcase10                 & 0.066    & 0.53    & 19.05      & 0.114    & 0.15    & 4.13       & 0.100    & 0.65    & -          & 0.492      & 0.02     & 2.53        & {\color[HTML]{333333} 0.088} & {\color[HTML]{333333} 0.34} & {\color[HTML]{333333} 4.40} \\
testcase13                 & 0.207    & 0       & 9.60       & 0.125    & 0.67    & 1.25       & 0.090    & 0       & -          & 0.665      & 0.40     & 0.61        & {\color[HTML]{333333} 0.121} & {\color[HTML]{333333} 0.68} & {\color[HTML]{333333} 2.21} \\
testcase14                 & 0.422    & 0       & 10.07      & 0.232    & 0.10    & 1.40       & 0.212    & 0       & -          & 0.864      & 0.03     & 0.87        & {\color[HTML]{333333} 0.220} & {\color[HTML]{333333} 0.43} & {\color[HTML]{333333} 1.43} \\
testcase15                 & 0.097    & 0.09    & 12.99      & 0.192    & 0       & 2.15       & 0.052    & 0.74    & -          & 0.549      & 0.07     & 1.04        & {\color[HTML]{333333} 0.066} & {\color[HTML]{333333} 0.76} & {\color[HTML]{333333} 2.49} \\
testcase16                 & 0.160    & 0.53    & 12.12      & 0.344    & 0.48    & 2.19       & 0.161    & 0.55    & -          & 0.971      & 0.26     & 0.98        & {\color[HTML]{333333} 0.163} & {\color[HTML]{333333} 0.40} & {\color[HTML]{333333} 2.32} \\
testcase19                 & 0.091    & 0.50    & 19.05      & 0.120    & 0.49    & 4.55       & -       & -       & -          & 0.367      & 0.02     & 2.50        & {\color[HTML]{333333} 0.047} & {\color[HTML]{333333} 0.70}  & {\color[HTML]{333333} 5.03} \\
testcase20                 & 0.118    & 0.71    & 18.75      & 0.107    & 0.74    & 4.58       & -       & -       & -          & 0.504      & 0.01     & 3.22        & {\color[HTML]{333333} 0.065} & {\color[HTML]{333333} 0.80} & {\color[HTML]{333333} 5.20} \\ \hline
\hline

Avg.                       & 0.135    & 0.46    & 14.77      & 0.150    & 0.45    & 3.04       & 0.107    & 0.52    & -          & 0.588    & 0.12   & \textbf{1.68}   & \textbf{0.101}               & \textbf{0.62}               & 3.37                        \\
Norm.                      & 1.337    & 0.74    & 4.38       & 1.485    & 0.73    & 0.90       & 1.059    & 0.84    & -          & 5.821    & 0.18   & \textbf{0.49}   & \textbf{1.000}               & \textbf{1.00}               & 1.00                        \\ \hline
\end{tabular}
\label{table:experimental_results}
\end{table*}

In Table \ref{table:experimental_results}, we compare our algorithm with the winning teams of the ICCAD 2023 contest and the state-of-the-art method, IREDGe \cite{DBLP:conf/aspdac/ChhabriaAPPJS21}. Notably, IREDGe is a Pytorch reimplementation by us since its data input is unsuitable for this dataset. The third-place submission is a Python script, so they don't have the runtime results. Moreover, their codes crashed in the last two cases. The experimental results show that our algorithm achieves a 34\% reduction in MAE and a 26\% improvement in F1 compared to the first place. Compared to the team (third place) with the best average performance, our algorithm achieves a 6\% reduction in MAE and a 16\% improvement in F1. It is worth noting that the third-place team augments their training data by generating additional cases through parameter modifications (in total, approximately 5400 cases), which provides their model with a competitive advantage. For MAE, we obtain optimal results for 2 cases and suboptimal results for 5 cases, with 2 cases (testcase14 and testcase16) differing from the best results by less than $0.008$ mV. Regarding F1, our algorithm achieves the best results in 6 cases. High F1 scores indicate the superior capability of our algorithm in predicting the worst-case IR drop. The results for testcase14 exhibit subpar performance, possibly attributed to its narrow hotspot region. The IR drop map visualization of testcase14 is illustrated at the top of Figure \ref{fig:comapred_maps}(a). Due to the limited features, including only the current map, PDN density map, and effective distance map, as well as its simpler model, IREDGe exhibits very poor performance. However, its runtime is shorter since feature extraction is not required. In summary, our proposed algorithm achieves outstanding and robust performance without significant time overhead.
%TODO 解释IREGDe为什么差，强调F1是基于最差的IR drop

\subsection{Ablative Study}

\subsubsection{The errors of important IR drops in sign-off}
\begin{table}[htbp]
\setlength{\tabcolsep}{0.6 mm}
\caption{The errors of important IR drops in sign-off, where $W_{IR}$ indicates the worst-case IR drop, Max $IR_{err}$ denotes the maximum IR drop error, the percentages in parentheses denote a ratio relative to $W_{IR}$, and the units of them are mV.}
\centering
\begin{tabular}{|c|c|cc|cc|}
\hline
                           &                             & \multicolumn{2}{c|}{IREDGe \cite{DBLP:conf/aspdac/ChhabriaAPPJS21}} & \multicolumn{2}{c|}{Ours}                                                    \\ \cline{3-6} 
\multirow{-2}{*}{Testcase} & \multirow{-2}{*}{$W_{IR}$} & $W_{IR}$ error    & Max $IR_{err}$   & $W_{IR}$ error                       & Max $IR_{err}$                         \\ \hline
\hline

testcase7                  & 4.30                        & 0.059 ( 1.37\%)             & 2.62          & {\color[HTML]{333333} \textbf{0.038 ( 0.89\%)}} & {\color[HTML]{333333} \textbf{1.46}} \\
testcase8                  & 4.90                        & 0.131 ( 2.66\%)             & 2.53          & {\color[HTML]{333333} \textbf{0.015 ( 0.31\%)}} & {\color[HTML]{333333} \textbf{2.19}} \\
testcase9                  & 3.79                        & 1.023 (26.97\%)             & 2.77          & {\color[HTML]{333333} \textbf{0.175 ( 4.60\%)}} & {\color[HTML]{333333} \textbf{1.25}} \\
testcase10                 & 4.53                        & \textbf{1.155} (25.49\%)    & 2.62          & {\color[HTML]{333333} 1.179 (26.02.\%)}          & {\color[HTML]{333333} \textbf{2.56}} \\
testcase13                 & 10.57                       & 2.368 (22.41\%)             & 3.21          & {\color[HTML]{333333} \textbf{0.541 ( 5.12\%)}} & {\color[HTML]{333333} \textbf{2.18}} \\
testcase14                 & 13.15                       & 4.327 (32.91\%)             & 7.45          & {\color[HTML]{333333} \textbf{1.148 ( 8.73\%)}} & {\color[HTML]{333333} \textbf{5.04}} \\
testcase15                 & 5.78                        & 3.324 (57.50\%)             & 3.58          & {\color[HTML]{333333} \textbf{1.823 (31.53\%)}} & {\color[HTML]{333333} \textbf{2.98}} \\
testcase16                 & 7.57                        & 2.743 (36.26\%)             & 4.40          & {\color[HTML]{333333} \textbf{0.843 (11.14\%)}} & {\color[HTML]{333333} \textbf{3.45}} \\
testcase19                 & 1.72                        & 0.843 (48.96\%)             & 1.92          & {\color[HTML]{333333} \textbf{0.018 ( 1.07\%)}} & {\color[HTML]{333333} \textbf{0.86}} \\
testcase20                 & 2.43                        & 0.585 (24.12\%)             & 2.50          & {\color[HTML]{333333} \textbf{0.077 ( 3.16\%)}} & {\color[HTML]{333333} \textbf{0.80}} \\ \hline
\hline

Avg.                       & 5.87                        & 1.656 (28.19\%)             & 3.36          & \textbf{0.586 ( 9.97\%)}                        & \textbf{2.28}                        \\ \hline
\end{tabular}
\label{table:IR_drop_error_results}
\end{table}

Sign-off is an important stage in ensuring the robustness of the power grid. Consequently, it is a typical concern for designers to ensure that the worst-case IR drop remains within a predetermined threshold. Hence, Table \ref{table:IR_drop_error_results} presents the comparison of the worst-case IR drop error and the maximum IR drop error between our proposed algorithm and IREDGe. Table \ref{table:IR_drop_error_results} shows that the average worst-case IR drop error of our algorithm is 9.97\% of the average worst-case IR drop of all designs, marking a reduction of 18.22\% compared to IREDGe. Furthermore, half of the designs exhibit an error of less than 5\%. Regarding the maximum IR drop error, our algorithm outperforms IREDGe across all designs, with the average maximum IR drop error reduced by 32.14\% compared to IREDGe. Considering the majority of the MAX $IR_{err}$ values of our algorithm are clustered around approximately 2mV and static IR drop constraints range from 1\% to 2.5\% of VDD, the errors of our algorithm are acceptable in light of rapid computational performance. The experimental results further demonstrate the efficacy of our algorithm in accurately predicting the worst-case IR drop.

\subsubsection{The runtime analysis of our algorithm}
\begin{table}[htbp]
\setlength{\tabcolsep}{1.0 mm}
\caption{The runtime analysis of our algorithm, where Parse means the runtime of parsing SPICE netlists, $F_i$ indicates the runtime of extracting instance-level features, and $F_P$ denotes the runtime of extracting PDN-structure-level features.}
\centering
\begin{tabular}{|c|c|c|ccccc|}
\hline
                           & \# Nodes  & Design Area & \multicolumn{5}{c|}{Time(s)}                                                                 \\ \cline{4-8} 
\multirow{-2}{*}{Testcase} & (million) & ($mm^2$)       & Parse & $F_i$ & $F_P$ & Inference                   & Total                       \\ \hline
\hline

testcase7                  & 0.09      & 1.44        & 0.44  & 0.06       & 0.04        & {\color[HTML]{333333} 2.79} & {\color[HTML]{333333} 3.33} \\
testcase8                  & 0.08      & 1.44        & 0.39  & 0.06       & 0.03        & {\color[HTML]{333333} 2.60} & {\color[HTML]{333333} 3.08} \\
testcase9                  & 0.17      & 2.77        & 0.84  & 0.10       & 0.05        & {\color[HTML]{333333} 3.20} & {\color[HTML]{333333} 4.19} \\
testcase10                 & 0.16      & 2.77        & 0.85  & 0.11       & 0.06        & {\color[HTML]{333333} 3.38} & {\color[HTML]{333333} 4.40} \\
testcase13                 & 0.02      & 0.26        & 0.18  & 0.01       & 0.01        & {\color[HTML]{333333} 2.01} & {\color[HTML]{333333} 2.21} \\
testcase14                 & 0.02      & 0.26        & 0.12  & 0.01       & 0.01        & {\color[HTML]{333333} 1.29} & {\color[HTML]{333333} 1.43} \\
testcase15                 & 0.06      & 0.95        & 0.43  & 0.05       & 0.04        & {\color[HTML]{333333} 1.98} & {\color[HTML]{333333} 2.49} \\
testcase16                 & 0.06      & 0.95        & 0.36  & 0.04       & 0.02        & {\color[HTML]{333333} 1.90} & {\color[HTML]{333333} 2.32} \\
testcase19                 & 0.18      & 3.02        & 0.96  & 0.11       & 0.06        & {\color[HTML]{333333} 3.90} & {\color[HTML]{333333} 5.03} \\
testcase20                 & 0.17      & 3.02        & 0.89  & 0.10       & 0.06        & {\color[HTML]{333333} 4.15} & {\color[HTML]{333333} 5.20} \\ \hline
\hline

Avg.                       & 0.10      & 1.69        & 0.55  & 0.06       & 0.04        & 2.72                        & 3.37                        \\ \hline
\end{tabular}
\label{table:runtime_analysis_results}
\end{table}

Table \ref{table:runtime_analysis_results} is a temporal analysis of all phases of our algorithm. Observably, as the technology node and design area increase, the predominant temporal increase of feature extraction is attributed to the time expended in parsing SPICE netlists. In Section \ref{sec:features}, we have demonstrated that the time complexity of our instance-level feature extraction methodology is $O\left(9\times K\times I\right)$, while that for PDN-structure-level feature extraction is $O(WN)$, indicating neither contribute excessively to the overall runtime. The experimental results presented in Table \ref{table:runtime_analysis_results} further demonstrate the minimal runtime required for feature computation. Although, to acquire the current, effective distance and PDN density maps, parsing SPICE netlists is also necessary, the 2023 ICCAD contest has provided these three maps, thereby exempting IREDGe from the feature extraction. Consequently, IREDGe exhibits the shortest runtime. However, in practical design scenarios, designers need to generate these maps independently. Hence, we believe that, by integrating the extraction of our new features with the extraction of these three maps, the runtime of our algorithm would be comparable with IREDGe.

\subsubsection{Impact of various enhancements}
\begin{figure}[h]
    \centering
    \includegraphics[width=1\linewidth]{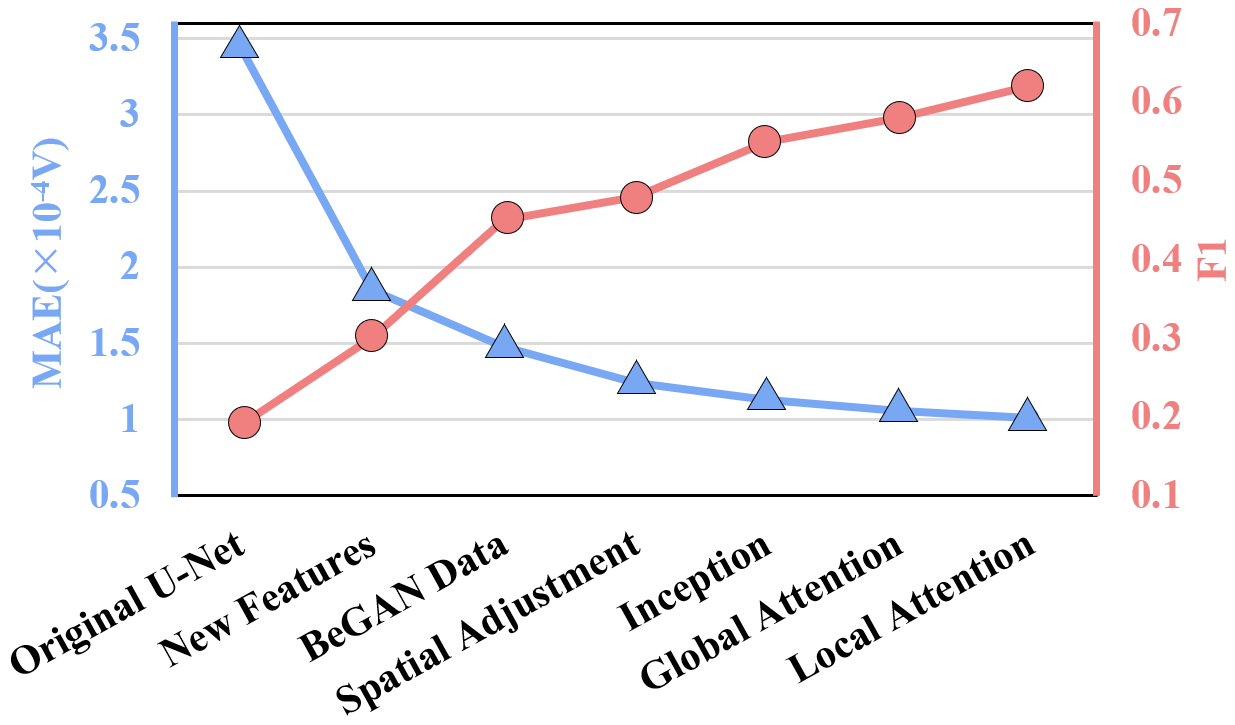}
    \caption{The changes in MAE and F1 with different enhancements in our model and training data, where the enhancements are incrementally added.}
    \label{fig:various_improvements}
\end{figure}

The variations of MAE and F1 with different enhancements in our model and training data are depicted in Figure \ref{fig:various_improvements}. From the two line charts, it can be observed that adding new features (Figure \ref{fig:features_and_label}(d)(e)(f)(g)) and additional data contribute significantly to performance improvement. Except that, our proposed spatial adjustment techniques have a notable impact on MAE, possibly due to our spatial adjustment techniques increasing the sampling probability of the case boundaries. Replacing traditional convolution with Inception has a greater contribution to improving F1, possibly due to Inception's capability to determine whether a point is a hotspot by leveraging information at different scales. Furthermore, our proposed global and local attention mechanisms, offering richer information to the model, contribute to the enhancement of model performance.

\subsubsection{Comparison with the results of IREDGe}
\begin{figure}[h]
    \centering
    \includegraphics[width=1\linewidth]{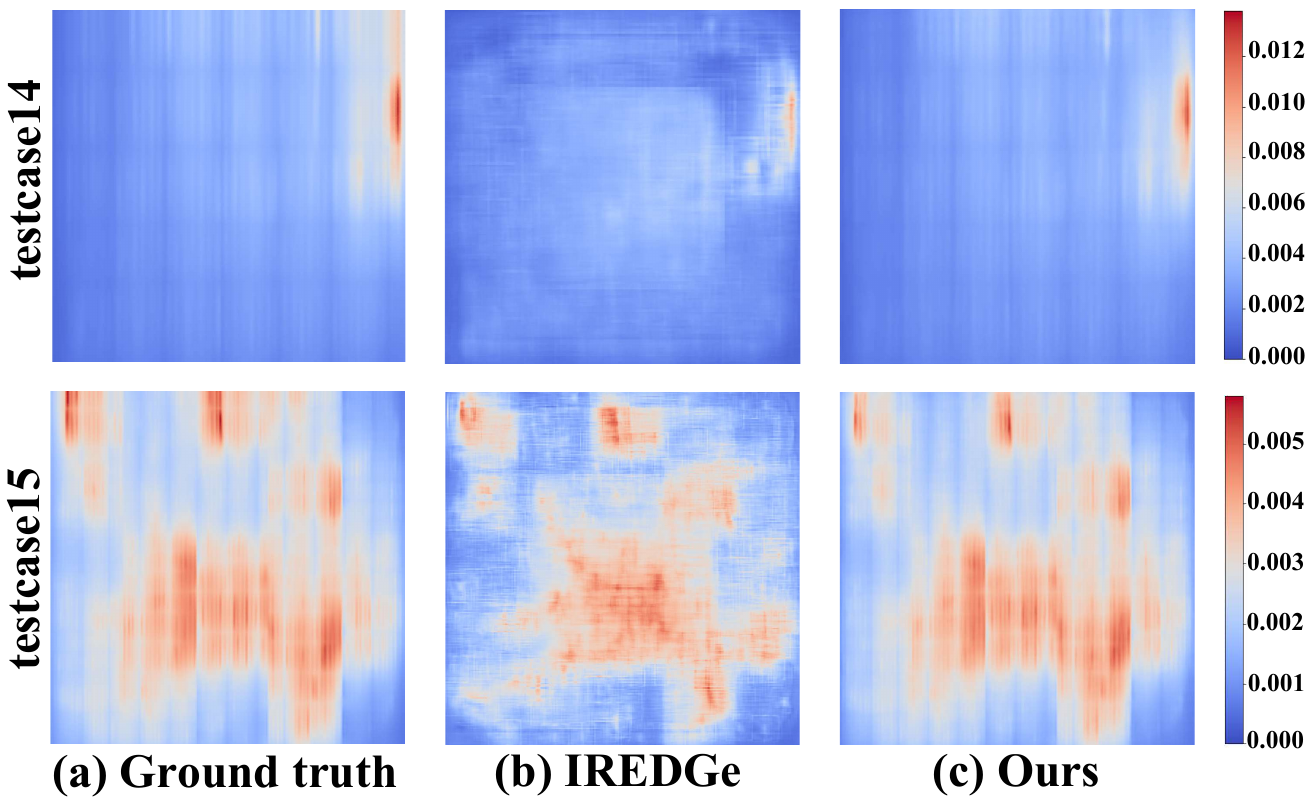}
    \caption{IR drop comparisons with IREDGe.}
    \label{fig:comapred_maps} 
\end{figure}
The comparisons of IR drop maps are illustrated in Figure \ref{fig:comapred_maps}. We present two cases, with testcase14 in the upper row and testcase15 in the lower row. The figure illustrates three distinct sets of IR drop maps: (a) the ground truth, (b) the IREDGe predicted maps, and (c) the maps generated by our algorithm. Due to the fact that IREDGe utilizes fewer features, it struggles to depict the boundaries of hotspots accurately. Specifically, IREDGe also fails to identify the stripe where resistances are located. However, the maps predicted by our algorithm closely match the ground truth maps.

\section{Conclusion}
\label{sec:conclusion} %TODO 提出的这些新技术可以干嘛
In this paper, we proposed a global and local attention-based Inception U-Net for static IR drop prediction. Moreover, we proposed several novel features and effective spatial adjustment techniques. Experimental results demonstrate that our algorithm can achieve the best performance compared to the winning teams of the ICCAD 2023 contest and the state-of-the-art method. In future work, we will explore the utilization of graph neural networks (GNNs) to better leverage the PDN structure.

\footnotesize
\bibliographystyle{unsrt}
\bibliography{references}

\end{document}